\documentclass{elsart}
\usepackage{amsfonts}

\usepackage{graphicx}
\usepackage{amsmath}

\begin{document}

\begin{frontmatter}
\title{On the meaning and interpretation of Tomography in abstract Hilbert spaces}
\author[rus]{V.I. Man'ko\corauthref{cor1}},
\ead{manko@na.infn.it}
\author[na]{G. Marmo}, \ead{marmo@na.infn.it}
\author[na]{A. Simoni\corauthref{cor1}},
\corauth[cor1]{Corresponding authors} \ead{simoni@na.infn.it}
\author[bama]{A. Stern}, \ead{astern@bama.ua.edu}
\author[george]{E.C.G. Sudarshan},
\author[na]{F. Ventriglia}\ead{ventriglia@na.infn.it}
\address[rus]{ P.N.Lebedev Physical Institute, Leninskii Prospect 53, Moscow 119991, Russia}
\address[na]{Dip. Sc. Fisiche dell'Universit\`{a} Federico II e Sez. INFN di Napoli, \\ Compl. Univ. Monte S.Angelo, I-80126 Naples, Italy}
\address[bama]{Department of Physics, University of Alabama, Tuscaloosa, AL 35487, USA}
\address[george]{Department of Physics, University of Texas, Austin, Texas 78712, USA}

\begin{abstract}

The mechanism of describing quantum states by standard probability
(tomographic one) instead of wave function or density matrix is
elucidated. Quantum tomography is formulated in an abstract Hilbert
space framework, by means of the identity decompositions in the
Hilbert space of hermitian linear operators, with trace formula as
scalar product of operators. Decompositions of identity are
considered with respect to over-complete families of projectors
labeled by extra parameters and containing a measure, depending on
these parameters. It plays the role of a Gram-Schmidt
orthonormalization kernel. When the measure is equal to one, the
decomposition of identity coincides with a positive operator valued
measure (POVM)  decomposition. Examples of spin tomography, photon
number tomography and symplectic tomography are reconsidered in this
new framework.
\end{abstract}
\end{frontmatter}

\section{Introduction}

Quantum states are described by vectors in a Hilbert space \cite{Dirac}, or
wave functions \cite{Schrod1926}, in the case of pure states. In the case of
mixed states, the density operators \cite{Landau27,vonNeum27} are used
instead to describe quantum states. On the other hand, different other tools
have been introduced to describe quantum states by means of functions on
phase space. These functions, like Wigner function \cite{Wig32}, Husimi-Kano
Q-quasidistribution \cite{Hus40,Kano56}, Sudarshan-Glauber diagonal
(singular) P-quasidistribution \cite{Sud63,Glau63}, contain informations
about the quantum state which amount to the informations carried by the
density matrix in an arbitrary representation. In fact, these different
quasidistributions are alternative, essentially equivalent forms of
representing the density operators. These quasidistributions have some
properties similar to those of classical probability distributions on phase
space, but they are not fair joint probability distributions since the
uncertainty relation of position and momentum is incompatible with the
existence of such probability density.

Recently a tomographic approach to reconstruct Wigner functions from optical
tomograms was suggested \cite{Ber-Bei,Vog-Ris}. The optical tomography
approach was generalized to provide symplectic tomography \cite
{Mancini95,d'Ariano96}. In the tomographic approach the quantum state is
associated to a probability distribution depending on some extra parameters.
This observation was used to develop a probability representation of quantum
mechanics in which the tomographic probability distribution (tomogram) is
considered as the primary object obeying an evolution equation of
generalized Fokker-Planck type \cite{Mancini97PL-Found Phys} and containing
all informations on quantum state. Thus it is possible to formulate quantum
mechanics by describing a quantum state by fair probability distribution
instead of wave functions or density matrices. By reading this chain of
associations backwards, it is quite natural to ask if it is possible to
provide an interpretation of tomograms directly at the level of the abstract
Hilbert space. In particular this interpretation should work equally well
for finite level systems (spin tomography) and generic systems. The
interpretation can make obvious the mechanism of description of quantum
states by fair probabilities instead of wave functions and density matrices.

It is the aim of this paper to show how to provide such an interpretation
without, however, indulging on more technical aspects (these will be
considered elsewhere). The main idea consists of expressing the tomogram in
terms of a scalar product in the space $\mathbb{H}$ of rank-one projectors,
that is in the linear space of operators acting on the space of quantum
states. These projectors are connected with special families of vectors in
the Hilbert space $\mathcal{H}$ of quantum states. The vectors are
eigenvectors of families of operators depending on some extra parameters. We
will obtain a decomposition of the identity operator in terms of a weighted
sum of the projectors depending on the extra parameter and determining the
tomogram, so that any matrix and in particular the density matrix can be
obtained as a weighted linear combination of the basis vectors in the space $%
\mathbb{H}$. This explains why the inversion formula (reconstruction
formula) works for the tomographic maps. The tomograms can be constructed
also for spin states \cite{DodPL,OlgaJept}. The relation between tomographic
maps and star-product quantization schemes was clarified in Ref. \cite
{BeppeJPA,PhysScr}. We develop our theory bearing in mind the case of spin
tomography. However, that general picture of tomographic map is applicable
to other kinds of tomographies too, e.g. to photon number tomography \cite
{BanWog,WogW,TombesEPL} and symplectic tomography. To demonstrate the
results we review the approach in which an $n\times n-$matrix is considered
as an $n^{2}-$vector, used for instance in Ref. \cite{SudJ.PL-Phys.Lett}.

The paper is organized as follows. In the next section 2 we review the
picture where a matrix is regarded as a vector. In section 3 we define the
tomography in abstract finite dimensional Hilbert spaces and give new
interpretation to the tomograms, both in terms of sets of rank-one
projectors and of \ families of unitary, or Hermitian, operators. In section
4 we discuss in the light of our picture some known examples as the spin
tomography for the finite dimensional case, the photon number and the
symplectic tomographies for the infinite dimensional case, deriving new
identity decompositions in terms of rank-one projectors in Hilbert space of
bounded Hermitian operators. Conclusions and perspectives are discussed in
section 5.

\section{Matrices as vectors}

In order to make clear how the tomographic approach provides relations
connecting probability distribution with density matrix elements we start
with a very elementary example. We consider two Hilbert spaces $\mathcal{H}$
and $\mathbb{H}.$ For simplicity we first identify the Hilbert spaces $%
\mathcal{H}$ with the qu-bit (i.e. spin 1/2) quantum state set, i.e. with
vectors
\begin{equation}
\left| \psi \right\rangle =\left[
\begin{array}{c}
\psi _{1} \\
\psi _{2}
\end{array}
\right] .
\end{equation}
Then the density matrix (density operator) for this pure state has the form
\begin{equation}
\hat{\rho}_{\psi }=\left| \psi \right\rangle \left\langle \psi \right| =%
\left[
\begin{array}{cc}
\psi _{1}\psi _{1}^{\ast } & \psi _{1}\psi _{2}^{\ast } \\
\psi _{2}\psi _{1}^{\ast } & \psi _{2}\psi _{2}^{\ast }
\end{array}
\right] .  \label{matrix}
\end{equation}
It is well known (e.g., see \cite{SudJKLKBregenz}) that the set of operators
acting on the Hilbert space $\mathcal{H}$ is a linear space. This space is a
Hilbert space $\mathbb{H}$ since one has the scalar product of two operators
$\hat{A}$ and $\hat{B}$ acting on the space $\mathcal{H}$ given by the
formula
\begin{equation}
\left\langle \hat{A}|\hat{B}\right\rangle =\mathrm{Tr}(\hat{A}^{\dagger }%
\hat{B}).  \label{scal prod}
\end{equation}
We use here the Dirac's notation for the scalar product. In fact, to write
an operator (a matrix) as a vector, being very simple, is convenient. Thus
the matrix of eq.(\ref{matrix}) can be mapped onto a 4-vector using the rule
\begin{equation}
\hat{\rho}_{\psi }\longrightarrow \left| \rho _{\psi }\right\rangle =\left[
\begin{array}{c}
\psi _{1}\psi _{1}^{\ast } \\
\psi _{1}\psi _{2}^{\ast } \\
\psi _{2}\psi _{1}^{\ast } \\
\psi _{2}\psi _{2}^{\ast }
\end{array}
\right]
\end{equation}
This rule allows to reconstruct a matrix if the corresponding 4-vector is
given. For example, given a 4-vector $\left| A\right\rangle $, one obtains
the matrix $\hat{A}$ as:
\begin{equation}
\left| A\right\rangle =\left[
\begin{array}{c}
a_{1} \\
a_{2} \\
a_{3} \\
a_{4}
\end{array}
\right] \longrightarrow \hat{A}=\left[
\begin{array}{cc}
a_{1} & a_{2} \\
a_{3} & a_{4}
\end{array}
\right]  \label{vecmatrix}
\end{equation}

Of course other "reconstructions" would be possible, and indeed this
possibility has been exploited to consider alternative associative products
on the vector space of matrices.\cite{BM}

The scalar product of eq.(\ref{scal prod}) expressed in terms of matrices $%
\hat{A}$ and $\hat{B}$ is nothing but the standard vector scalar product
given by
\begin{equation}
\left\langle \hat{A}|\hat{B}\right\rangle =\left\langle A|B\right\rangle
=\sum_{k=1}^{4}a_{k}^{\ast }b_{k}
\end{equation}
The set of 4-vectors equipped with this scalar product is the Hilbert space $%
\mathbb{H.}$ Thus, having an initial Hilbert space of vectors $\mathcal{H}$
of two-dimensional qu-bit, we have also the four-dimensional Hilbert space
of 4-vectors $\mathbb{H}=B(\mathcal{H}),$ the linear operators acting on the
Hilbert space $\mathcal{H.}$ The orthogonal basis in the space $\mathcal{H}$
of spin up and down states (standard basis)
\begin{equation}
\left| e_{1}\right\rangle =\left[
\begin{array}{c}
1 \\
0
\end{array}
\right] ;\left| e_{2}\right\rangle =\left[
\begin{array}{c}
0 \\
1
\end{array}
\right] ,
\end{equation}
for instance, is complete. The completeness relation can be given in the
form of an equality valid in the Hilbert space $\mathbb{H}=B(\mathcal{H}),$
namely
\begin{equation}
\hat{P}_{1}+\hat{P}_{2}=\mathbb{I}_{\mathcal{H}}.  \label{projecomplet}
\end{equation}
Here the orthogonal projectors
\begin{equation}
\hat{P}_{1}=\left| e_{1}\right\rangle \left\langle e_{1}\right| =\left[
\begin{array}{cc}
1 & 0 \\
0 & 0
\end{array}
\right] ;\hat{P}_{2}=\left| e_{2}\right\rangle \left\langle e_{2}\right| =%
\left[
\begin{array}{cc}
0 & 0 \\
0 & 1
\end{array}
\right]
\end{equation}
satisfying the relation (\ref{projecomplet}) corresponds to the possibility
to decompose each vector of $\mathcal{H}$ in terms of the basis vector $%
\left| e_{1}\right\rangle $ and $\left| e_{2}\right\rangle ,$ that is
\begin{equation}
\left| \psi \right\rangle =\psi _{1}\left| e_{1}\right\rangle +\psi
_{2}\left| e_{2}\right\rangle .
\end{equation}
We considered in such details very simple properties of the linear spaces $%
\mathcal{H}$ and $\mathbb{H}$, because essentially they are the
basis of our considerations on the possibility to construct
tomographic probabilities and to guarantee the existence of an
inversion formula yielding the operator from its tomogram. In
general, the completeness relation of a basis $\left| \mu
\right\rangle $ of $\mathcal{H}$ can be represented in the form
\begin{equation}
\sum\nolimits_{\mu }\hat{P}_{\mu }=\mathbb{I}_{\mathcal{H}}.
\end{equation}
The rank-one projectors
\begin{equation}
\hat{P}_{\mu }=\left| \mu \right\rangle \left\langle \mu \right|
\end{equation}
depend on a set of parameters $\mu $ (discrete or continuous, as well as
finite or infinite) and, being in general non-orthogonal:
\begin{equation*}
\hat{P}_{\mu }\hat{P}_{\mu ^{\prime }}\neq 0,
\end{equation*}
they form a positive operator valued measure (POVM).

On the other hand, in the four-dimensional space $\mathbb{H}$ one has the
standard basis
\begin{equation}
\left| B_{1}\right\rangle =\left[
\begin{array}{c}
1 \\
0 \\
0 \\
0
\end{array}
\right] ,\left| B_{2}\right\rangle =\left[
\begin{array}{c}
0 \\
1 \\
0 \\
0
\end{array}
\right] ,\left| B_{3}\right\rangle =\left[
\begin{array}{c}
0 \\
0 \\
1 \\
0
\end{array}
\right] ,\left| B_{4}\right\rangle =\left[
\begin{array}{c}
0 \\
0 \\
0 \\
1
\end{array}
\right] ,
\end{equation}
so that each vector $\left| A\right\rangle $ may be decomposed as
\begin{equation}
\left| A\right\rangle =\sum_{k=1}^{4}a_{k}\left| B_{k}\right\rangle .
\end{equation}
This decomposition of the vector $\left| A\right\rangle $ corresponds to the
decomposition of the matrix $\hat{A}$ of eq.(\ref{vecmatrix}) in the form
\begin{equation*}
\left[
\begin{array}{cc}
a_{1} & a_{2} \\
a_{3} & a_{4}
\end{array}
\right] =a_{1}\left[
\begin{array}{cc}
1 & 0 \\
0 & 0
\end{array}
\right] +a_{2}\left[
\begin{array}{cc}
0 & 1 \\
0 & 0
\end{array}
\right] +a_{3}\left[
\begin{array}{cc}
0 & 0 \\
1 & 0
\end{array}
\right] +a_{4}\left[
\begin{array}{cc}
0 & 0 \\
0 & 1
\end{array}
\right] ,
\end{equation*}
with respect to a basis of non-Hermitian operators. However, it is
always possible to use a basis of Hermitian operators, for instance
the unit
matrix $\hat{\sigma}_{0}$ together with the Pauli matrices $\hat{\sigma}%
_{k},(k=1,2,3)$. Then\ :
\begin{equation}
\left| \sigma _{0}\right\rangle =\frac{1}{2}\left[
\begin{array}{c}
1 \\
0 \\
0 \\
1
\end{array}
\right] ,\left| \sigma _{1}\right\rangle =\frac{1}{2}\left[
\begin{array}{c}
0 \\
1 \\
1 \\
0
\end{array}
\right] ,\left| \sigma _{2}\right\rangle =\frac{1}{2}\left[
\begin{array}{c}
0 \\
-i \\
i \\
0
\end{array}
\right] ,\left| \sigma _{3}\right\rangle =\frac{1}{2}\left[
\begin{array}{c}
1 \\
0 \\
0 \\
-1
\end{array}
\right] ,
\end{equation}
so that
\begin{equation}
\left| A\right\rangle =\sum_{k=0}^{3}\alpha _{k}\left| \sigma
_{k}\right\rangle .
\end{equation}
The relation between old and new components is
\begin{equation*}
\left[
\begin{array}{cc}
\alpha _{0} & \alpha _{1} \\
\alpha _{2} & \alpha _{3}
\end{array}
\right] =\left[
\begin{array}{cc}
a_{1}+a_{4} & a_{2}+a_{3} \\
i\left( a_{2}-a_{3}\right)  & a_{1}-a_{4}
\end{array}
\right] ;\left[
\begin{array}{cc}
a_{1} & a_{2} \\
a_{3} & a_{4}
\end{array}
\right] =\frac{1}{2}\left[
\begin{array}{cc}
\alpha _{0}+\alpha _{3} & \alpha _{1}-i\alpha _{2} \\
\alpha _{1}+i\alpha _{2} & \alpha _{0}-\alpha _{3}
\end{array}
\right] .
\end{equation*}

So, in general, if we were to use the coefficient of expansion in
terms of a Hermitian set of $n$ basis matrices, vectors
corresponding to Hermitian operators would be real and the Hermitian
conjugation would act on them as the identity. In the case of
non-Hermitian basis, those vectors are not real, in general, and
belong to a real linear $n^{2}-$dimensional submanifold of
$\mathbb{C}^{n^{2}}$  which is invariant under Hermitian
conjugation.

Now we transfer our discussion of the completeness relation of a basis of
vectors, given for the two-dimensional Hilbert space $\mathcal{H}$, to the
case of the four-dimensional Hilbert space $\mathbb{H}$. Thus we get
4-projectors
\begin{equation}
\mathbb{P}_{k}=\left| B_{k}\right\rangle \left\langle B_{k}\right| ,\quad
k=1,...,4.
\end{equation}
The completeness relation in $\mathbb{H}$ reads
\begin{equation}
\sum\limits_{k=1}^{4}\mathbb{P}_{k}=\mathbb{I}_{\mathbb{H}},  \label{baspro}
\end{equation}
where $\mathbb{I}_{\mathbb{H}}$\ is the four-dimensional unit
matrix.
Alternatively, in the four-dimensional space, one can have a POVM $\mathbb{P}%
_{\mu }$\ with a set of parameters such that
\begin{equation}
\sum\nolimits_{\mu }\mathbb{P}_{\mu }=\mathbb{I}_{\mathbb{H}}.
\label{decom4}
\end{equation}
This property means that one can decompose any $2\times 2$ matrix in terms
of the $2\times 2$ matrices corresponding to the projectors $\mathbb{P}_{\mu
}$\ . In matrix form the previous eq.(\ref{decom4}) reads
\begin{equation}
\sum\nolimits_{\mu }\left( \mathbb{P}_{\mu }\right) _{ij,mn}=\delta
_{im}\delta _{jn},\quad i,j,m,n=1,2.  \label{famidec}
\end{equation}
That means that the index $k$ in eq.(\ref{baspro}) is considered a double
index in labelling the elements of two by two matrices. In principle one can
have more complicated conditions of completeness when the projectors $%
\mathbb{P}_{\mu }$ are not orthogonal and have trace different from unity.
Then a Gram-Schmidt orthogonalization procedure can be encoded by an extra
kernel in the relation (\ref{famidec}). We will see that precisely this
situation takes place in some of the examples considered later on.

\section{The abstract Hilbert space definition of tomograms}

\subsection{Tomographics sets}

In general, any kind of tomogram of a pure state $\left| \psi \right\rangle $
is the positive real number $\mathcal{W}_{\psi }(\alpha ,\beta ,...),$
depending on a set of parameters $(\alpha ,\beta ,...)$ which label a set of
states $\left| \alpha ,\beta ,...\right\rangle ,$ defined as:
\begin{equation}
\mathcal{W}_{\psi }(\alpha ,\beta ,...)=\left| \left\langle \alpha ,\beta
,...|\psi \right\rangle \right| ^{2}.
\end{equation}
At a first glance, it seems quite difficult to read the tomogram as a scalar
product, rather than a square modulus. Nevertheless, this is possible by
thinking in terms of rank-one projectors rather than of (pure) states. In
fact, by using the density operator $\hat{\rho}=\left| \psi \right\rangle
\left\langle \psi \right| $ and the projectors $P_{\alpha ,\beta
,...}=\left| \alpha ,\beta ,...\right\rangle \left\langle \alpha ,\beta
,...\right| ,$ we may interpret the tomogram as a scalar product on the
space of rank-one projectors:
\begin{equation}
\mathcal{W}_{\psi }(\alpha ,\beta ,...)=\left| \left\langle \alpha ,\beta
,...|\psi \right\rangle \right| ^{2}=\mathrm{Tr}\left( \hat{\rho}P_{\alpha
,\beta ,...}\right) .
\end{equation}
This definition may be applied also in the case of an arbitrary density
operator $\hat{\rho}.$

In the following we wish to characterize the sets of vectors $\left| \alpha
,\beta ,...\right\rangle $ which allow for a complete reconstruction of the
state $\left| \psi \right\rangle $, or an arbitrary density operator $\hat{%
\rho},$ from the knowledge of its tomograms $\mathcal{W}_{\psi }(\alpha
,\beta ,...).$ These sets will be called \textsl{tomographic sets}. In the
light of our interpretation of the tomogram, the meaning of such a
reconstruction is nothing but a consequence of a decomposition of identity
in the space of rank-one projectors in terms of the family $\left| P_{\alpha
,\beta ,...}\right\rangle \left\langle P_{\alpha ,\beta ,...}\right|,$ after
taking into account that the projectors $P_{\alpha ,\beta ,...}$ in general
are not orthogonal. We will discuss our interpretation in the finite
dimensional Hilbert spaces. Our construction essentially goes along the
following lines.

Suppose a set $\{\left| e_{\alpha \beta }\right\rangle \}_{\alpha ,\beta
=1}^{n}$\ of $n^{2}$\ vectors of $C^{n}$\ is found in such a way that the
respective projectors $\left| e_{\alpha \beta }\right\rangle \left\langle
e_{\alpha \beta }\right| $\ are a basis $\{\left| P_{k}\right\rangle
\}_{k=1}^{n^{2}}$\ of$\ C^{n^{2}}=C^{n}\otimes C^{n}=B(C^{n}).$\emph{\ }We
use here a collective index $k$ instead of $(\alpha ,\beta )$, e.g. $%
k=(\alpha -1)n+\beta $. By means of the Gram-Schmidt procedure, for
instance, we may convert the basis $\{\left| P_{k}\right\rangle
\}_{k=1}^{n^{2}}$\ into an orthonormal basis $\{\left| V_{j}\right\rangle
\}_{j=1}^{n^{2}}:$
\begin{equation}
\left| V_{j}\right\rangle =\sum\limits_{k=1}^{n^{2}}\gamma _{jk}\left|
P_{k}\right\rangle \quad ,\quad \left\langle V_{i}|V_{j}\right\rangle
=\delta _{ij}.
\end{equation}
In general, every element of the orthonormal basis $\{\left|
V_{j}\right\rangle \}$ is a linear combination of projectors, rather than a
single projector like $\left| P_{k}\right\rangle $\ associated to a vector
of $\mathbb{C}^{n}.$

There exists a decomposition of the identity on $\mathbb{C}^{n^{2}}=B(%
\mathbb{C}^{n})$%
\begin{equation}
\mathbb{I}_{n^{2}}=\sum\limits_{j=1}^{n^{2}}\left| V_{j}\right\rangle
\left\langle V_{j}\right| =\sum\limits_{j,k,l=1}^{n^{2}}\gamma _{jk}^{\ast
}\gamma _{jl}\hat{P}_{l}\mathrm{Tr}(\hat{P}_{k}\cdot
)=\sum\limits_{l=1}^{n^{2}}\hat{K}_{l}\mathrm{Tr}(\hat{P}_{l}\cdot ),
\label{identity}
\end{equation}
where the Gram-Schmidt kernel $\hat{K}_{l}$ has been introduced
\begin{equation}
\hat{K}_{l}=\sum\limits_{j,k=1}^{n^{2}}\gamma _{jl}^{\ast }\gamma _{jk}\hat{P%
}_{k}.  \label{GS}
\end{equation}
We observe that $\hat{K}_{l}$ is a nonlinear function of the projectors $%
\hat{P}_{k}$, because also the coefficients $\gamma$'s depend on the
projectors.

We define the set $\{\left| e_{\alpha \beta }\right\rangle \}_{\alpha ,\beta
=1}^{n}$ of $n^{2}$ vectors of $\mathbb{C}^{n}$ a \textsl{minimal
tomographic set}. The tomogram of a density matrix $\hat{\rho}$ with respect
to this minimal tomographic set is defined by
\begin{equation}
\mathcal{W}_{\rho }(\alpha ,\beta )=\mathrm{Tr}(\left| e_{\alpha \beta
}\right\rangle \left\langle e_{\alpha \beta }\right| \hat{\rho}),\quad
(\alpha ,\beta =1,...,n).
\end{equation}
Then, from the decomposition of identity in terms of the tomographic
projectors, we get an inversion formula for the density matrix $\hat{\rho}$
or any other operator on $\mathbb{C}^{n}:$%
\begin{equation}
\hat{\rho}=\sum\limits_{j,k,l=1}^{n^{2}}\gamma _{jk}^{\ast }\gamma _{jl}\hat{%
P}_{l}\mathrm{Tr}(\hat{P}_{k}\hat{\rho})=\sum\limits_{\mu ,\nu =1}^{n}\left[
\sum\limits_{j,k,l=1}^{n^{2}}\gamma _{jk}^{\ast }\gamma _{jl}(\hat{P}%
_{k})_{\mu \nu }^{\ast }\hat{P}_{l}\right] (\hat{\rho})_{\mu \nu }.
\end{equation}
In other words, writing the previous equation in terms of matrix elements:
\begin{equation}
(\hat{\rho})_{\mu ^{\prime }\nu ^{\prime }}=\sum\limits_{\mu ,\nu =1}^{n}%
\left[ \sum\limits_{j,k,l=1}^{n^{2}}\gamma _{jk}^{\ast }\gamma _{jl}(\hat{P}%
_{k})_{\mu \nu }^{\ast }(\hat{P}_{l})_{\mu ^{\prime }\nu ^{\prime }}\right] (%
\hat{\rho})_{\mu \nu }
\end{equation}
we have the corresponding expression for the decomposition of the identity:
\begin{eqnarray}
\sum\limits_{j,k,l=1}^{n^{2}}\gamma _{jk}^{\ast }\gamma _{jl}(\hat{P}%
_{k})_{\mu \nu }^{\ast }(\hat{P}_{l})_{\mu ^{\prime }\nu ^{\prime }}
&=&\sum\limits_{k=1}^{n^{2}}(\hat{P}_{k})_{\mu \nu }^{\ast
}(\sum\limits_{j,l=1}^{n^{2}}\gamma _{jk}^{\ast }\gamma _{jl}\hat{P}%
_{l})_{\mu ^{\prime }\nu ^{\prime }}  \notag \\
&=&\delta _{\mu \mu ^{\prime }}\delta _{\nu \nu ^{\prime }}=\left( \mathbb{I}%
_{n}\otimes \mathbb{I}_{n}\right) _{\mu \mu ^{\prime },\nu \nu ^{\prime }}.
\end{eqnarray}

\noindent\ A set containing more than $n^{2}$ vectors of $\mathbb{C}^{n}$ is
a tomographic set when it contains a minimal set. In other words, a
tomographic set is such that any vector belongs to a (minimal) tomographic
subset of $n^{2}$ vectors. In particular, a set is tomographic if any subset
of $n^{2}$ vectors is a (minimal) tomographic set.

Now, a basis of $n^{2}$\ rank-one projectors can always be found. In fact,
an orthonormal basis of $B(\mathbb{C}^{n})$ containing $n^{2}$ Hermitian
operators is associated with the generators $\tau _{k}$ of the group $U(n)$,
multiplying each element by the imaginary unit $i$. Each generator, using
its spectral decomposition, can be written in terms of projectors. Moreover,
each projector can be expressed by means of rank-one projectors. So, from
the spectral decompositions of the generators of $\ U\left( n\right) ,$ we
may extract a basis of $n^{2}$ rank-one projectors.

Alternatively, using an orthonormal basis $\{\left|\alpha \right\rangle
\}_{\alpha =1}^{n}$ of $n$ vectors of $\mathbb{C}^{n}$, we may define in $B(%
\mathbb{C}^{n})$ an orthogonal basis of $n^{2}$ Hermitian operators given
by:
\begin{equation}
\left\{ \left( \left| \alpha \right\rangle \left\langle \beta \right|
+\left| \beta \right\rangle \left\langle \alpha \right| \right) ,\quad
i\left( \left| \alpha \right\rangle \left\langle \beta \right| -\left| \beta
\right\rangle \left\langle \alpha \right| \right) \right\} ,\quad (\alpha
,\beta =1,...,n).
\end{equation}
Then, from their spectral decompositions we may extract $\ $a basis of $%
n^{2} $ rank-one projectors.

For example, starting from a fiducial (orthonormal) basis of $\mathbb{C}%
^{2}, $ two different suitable unitary operators
\begin{equation*}
U_{\alpha }=\left|
\begin{array}{cc}
a_{\alpha } & b_{\alpha } \\
-b_{\alpha }^{\ast } & a_{\alpha }^{\ast }
\end{array}
\right| ;\ a_{\alpha }a_{\alpha }^{\ast }+b_{\alpha }b_{\alpha }^{\ast }=1;\
\left( \alpha =1,2\right)
\end{equation*}
are needed to generate other two different basis of $\mathbb{C}^{2}$ and
obtain a tomographic set of six vectors containing three different minimal
tomographic sets. In fact, starting from the standard basis of $\mathbb{C}%
^{2}$, the matrix of 4-vectors
\begin{equation*}
\left[
\begin{array}{cccccc}
1 & 0 & a_{1}a_{1}^{\ast } & b_{1}b_{1}^{\ast } & a_{2}a_{2}^{\ast } &
b_{2}b_{2}^{\ast } \\
0 & 0 & -a_{1}b_{1} & a_{1}b_{1} & -a_{2}b_{2} & a_{2}b_{2} \\
0 & 0 & -a_{1}^{\ast }b_{1}^{\ast } & a_{1}^{\ast }b_{1}^{\ast } &
-a_{2}^{\ast }b_{2}^{\ast } & a_{2}^{\ast }b_{2}^{\ast } \\
0 & 1 & b_{1}b_{1}^{\ast } & a_{1}a_{1}^{\ast } & b_{2}b_{2}^{\ast } &
a_{2}a_{2}^{\ast }
\end{array}
\right]
\end{equation*}
has maximal rank, when the two different operators satisfy some extra
condition such as, for instance,
\begin{equation}
\Im (a_{1}b_{1}a_{2}^{\ast }b_{2}^{\ast })=\det \left|
\begin{array}{cc}
\Im \left( a_{1}b_{1}\right) & \Re \left( a_{1}b_{1}\right) \\
\Im \left( a_{2}b_{2}\right) & \Re \left( a_{2}b_{2}\right)
\end{array}
\right| \neq 0.
\end{equation}
This condition shows that the two complex numbers $a_{1}b_{1}$ and $%
a_{2}b_{2}$ cannot be proportional on the reals or, equivalently, they have
different phases.

\bigskip

Let us characterize more closely the manifold of rank-one projectors in the
real space $\mathbb{R}^{n^{2}}$of Hermitian operators . Using the basis of
the generators $\tau _{k}$ of $U(n)$, with $\tau _{1}=\mathbb{I}$ and $%
\mathrm{Tr}\tau _{k}=0$ $\left( k=2,...,n^{2}\right) ,$ we may express any
Hermitian operator $A$ as
\begin{equation}
A=\sum\limits_{k=1}^{n^{2}}\alpha ^{k}\tau _{k}.
\end{equation}
The manifold of rank-one projectors is given by the vectors whose components
$\left\{ \alpha ^{k}\right\} \in \mathbb{R}^{n^{2}}$ fulfill the conditions $%
A^{2}=A$ , $\mathrm{Tr}A=1$ (which implies $\alpha ^{1}=1/n$). In the dual
space $u^{\ast }\left( n\right) $ of the Lie algebra of the generators $\tau
_{k},$ this manifold is an orbit $\mathcal{O}_{\mathcal{P}}$ of the
co-adjoint action of the group $U(n)$. As the set of rank-one projectors may
be identified with the projective space $\mathbb{P}\left( \mathbb{C}%
^{n}\right) $ with $2\left( n-1\right) $ real dimensions, the orbit $%
\mathcal{O}_{\mathcal{P}}$ is the only orbit with the same dimensions placed
in the plane $\alpha ^{1}=1/n.$ Among the orbits, $\mathcal{O}_{\mathcal{P}}$
is the one with lowest dimensionality, as the stability group of its points
is just $U(1)\times $ $U(n-1).$ For instance, for $n=2,$ the rank-one
projector set is the Bloch sphere $S^{2}:\ \left\{ \ \left( \alpha
^{2}\right) ^{2}+\left( \alpha ^{3}\right) ^{2}+\left( \alpha ^{4}\right)
^{2}=1/4\right\} $ placed in the plane $\alpha ^{1}=1/2$.

A minimal tomographic set therefore is a set of $n^{2}$ projectors $\left\{
P(m_{k})\right\} $, where $m_{k}\in \mathcal{O}_{\mathcal{P}}$, such that
the linear span of $\left\{ P(m_{k})\right\} $ is the entire $\mathbb{R}%
^{n^{2}}$, that is $\det \left\{ \left( \alpha _{k}\right) ^{l}\right\} \neq
0.$ Of course, $\mathcal{O}_{\mathcal{P}}$ is a maximal tomographic set.

\subsection{Families of operators generating tomographic sets}

An interesting question is how to find a way to construct
tomographic sets. This question may be answered in different but
equivalent ways. We consider some of them here and provide a few
well known examples to show how our proposal works.

The first way consists in taking a fiducial rank-one projector $P_{0}$ and
acting on it with a suitable family of (at least $n^{2})$ unitary operators $%
U_{\alpha }$, depending on some parameters $\alpha $. The family has to be
chosen in such a manner that the set of projectors
\begin{equation}
P_{\alpha }=U_{\alpha }P_{0}U_{\alpha }^{\dagger }
\end{equation}
results into a tomographic one. This is granted only if the family $%
U_{\alpha }$ is not contained in any proper subgroup of $U(n)$ or,
equivalently, if the group generated by the family $U_{\alpha }$ is $U(n)$.
This condition is also sufficient if, moreover, from the family of unitary
operators it is possible to extract, \emph{via }the Cayley map, for
instance, a basis for the Lie algebra $u(n)$. Then, a family which is
``skew'' in the group $U(n)$ is a suitable \emph{tomographic family of
unitary operators}.

Alternatively, it is possible to start with a fiducial Hermitian operator $%
A_{0}$ and to act on it with a ``skew'' family of unitary operators $%
U_{\alpha }$, generating a family of (iso-spectral)$\ $Hermitian operators
\begin{equation}
A_{\alpha }=U_{\alpha }A_{0}U_{\alpha }^{\dagger }.
\end{equation}
Choosing $\ A_{0}$ to be generic, i.e. with simple eigenvalues, the action
of $U_{\alpha }$ on the rank-one projectors associated with the eigenstates
of $\ A_{0}$ gives rise to a tomographic set of projectors. In other words,
we may obtain a tomographic set from a suitable family of Hermitian
operators.

Taking $U(n)$ as a tomographic family we obtain a (maximal) decomposition of
the identity, analogous to eq. (\ref{identity}). However, integrating all
projectors $P$ over the symplectic orbit $\mathcal{O}_{\mathcal{P}}$ and
using the volume $\Omega =\omega ^{n-1}$ constructed with the canonical
symplectic form $\omega ,$we have to use a Hermitian kernel $\hat{K}(m),$
which is an operator valued function of the point $m$ on the orbit $\mathcal{%
O}_{\mathcal{P}},$ that plays the same role of the Gram-Schmidt kernel in
the minimal case:
\begin{equation}
\mathbb{I}_{n^{2}}=\int\nolimits_{\mathcal{O}_{\mathcal{P}}}\hat{K}(m)%
\mathrm{Tr}(P(m)\cdot )\Omega .  \label{fullidentity}
\end{equation}
For instance, in the $U(2)$ case, we have
\begin{equation}
\hat{K}(\theta ,\phi )=\frac{1}{4\pi }\left[
\begin{array}{cc}
1+3\cos \theta & 3e^{-i\phi }\sin \theta \\
3e^{i\phi }\sin \theta & 1-3\cos \theta
\end{array}
\right]\label{decompspinker}
\end{equation}
so that, for any operator $A,$it results
\begin{equation}
A=\int_{0}^{2\pi }\int_{0}^{\pi }\hat{K}(\theta ,\phi )\mathrm{Tr}(P(\theta
,\phi )A)\sin \theta d\theta d\phi .  \label{decompspin}
\end{equation}

By using eq. (\ref{fullidentity}) in the scalar product of any pair of
operators $A,B,$ we obtain
\begin{eqnarray*}
\mathrm{Tr}(AB) &=&\mathrm{Tr}\left( A\int\nolimits_{\mathcal{O}_{\mathcal{P}%
}}\hat{K}(m)\mathrm{Tr}(P(m)B)\Omega \right) =\int\nolimits_{\mathcal{O}_{%
\mathcal{P}}}\mathrm{Tr}\left( \hat{K}(m)A\right) \mathrm{Tr}(P(m)B)\Omega \\
&=&\int\nolimits_{\mathcal{O}_{\mathcal{P}}}\mathrm{Tr}\left( \hat{K}%
(m)B\right) \mathrm{Tr}(P(m)A)\Omega =\mathrm{Tr}\left( A\int\nolimits_{%
\mathcal{O}_{\mathcal{P}}}P(m)\mathrm{Tr}(\hat{K}(m)B)\Omega \right) .
\end{eqnarray*}
So, at least in a ``weak sense'',
\begin{equation*}
\mathbb{I}_{n^{2}}=\int\nolimits_{\mathcal{O}_{\mathcal{P}}}P(m)\mathrm{Tr}(%
\hat{K}(m)\cdot )\Omega \Longrightarrow \hat{K}(m)=\int\nolimits_{\mathcal{O}%
_{\mathcal{P}}}P(m^{\prime })\mathrm{Tr}(\hat{K}(m^{\prime })\hat{K}%
(m))\Omega ^{\prime }.
\end{equation*}
Finally, substituting the previous expression of $\hat{K}(m)$ in eq. (\ref
{fullidentity}), we obtain
\begin{equation}
\mathbb{I}_{n^{2}}=\int\nolimits_{\mathcal{O}_{\mathcal{P}}}\left[
\int\nolimits_{\mathcal{O}_{\mathcal{P}}}\mathrm{Tr}(\hat{K}(m^{\prime })%
\hat{K}(m))P(m^{\prime })\Omega ^{\prime }\right] \mathrm{Tr}(P(m)\cdot
)\Omega
\end{equation}
in full analogy with eq. (\ref{identity}). Hence $\hat{K}(m)$ is just a
Gram-Schmidt kernel, at least in a ``weak sense''.

In general a tomographic family of unitary operators ranges between a
minimal family and a maximal family, and is representative of the whole
group $U(n)$, so that necessarily the commutant of the family must be the
identity. Then, a decomposition of unity must be available by integrating on
the space of parameters of the family with a suitable kernel $\hat{K}(m).$

\section{Examples}

\subsection{Spin tomography}

For a qu-bit (i.e. a particle of spin 1/2), a tomographic iso-spectral
two-parameters family of Hermitian operators is
\begin{equation}
A(\theta ,\phi ):=\left[
\begin{array}{cc}
\cos \theta & e^{-i\phi }\sin \theta \\
e^{i\phi }\sin \theta & -\cos \theta
\end{array}
\right]
\end{equation}
A given operator of the family corresponds to the component of the spin (up
to a factor $\hbar /2)$ in the direction
\begin{equation*}
\overrightarrow{n}=(\sin \theta \cos \phi ,\sin \theta \sin \phi ,\cos
\theta ).
\end{equation*}
In fact
\begin{equation}
\overrightarrow{n}\cdot \overrightarrow{\sigma }=A(\theta ,\phi )
\end{equation}
The spectrum of $A(\theta ,\phi )$ and corresponding orthonormal
eigenvectors are
\begin{equation}
m_{\pm }=\pm 1,\quad \left| m_{+}\theta \phi \right\rangle =\left[
\begin{array}{c}
e^{-i\phi /2}\cos \theta /2 \\
e^{i\phi /2}\sin \theta /2
\end{array}
\right] ,\quad \left| m_{-}\theta \phi \right\rangle =\left[
\begin{array}{c}
e^{-i\phi /2}\sin \theta /2 \\
-e^{i\phi /2}\cos \theta /2
\end{array}
\right]
\end{equation}
while the respective projectors are
\begin{equation*}
\left| m_{+}\theta \phi \right\rangle \left\langle m_{+}\theta \phi \right|
= \left[
\begin{array}{cc}
\cos ^{2}\frac{1}{2}\theta & e^{-i\phi }\cos \frac{1}{2}\theta \sin \frac{1}{%
2}\theta \\
e^{i\phi }\sin \frac{1}{2}\theta \cos \frac{1}{2}\theta & \sin ^{2}\frac{1}{2%
}\theta
\end{array}
\right]
\end{equation*}
and
\begin{equation*}
\left| m_{-}\theta \phi \right\rangle \left\langle m_{-}\theta \phi \right|
= \left[
\begin{array}{cc}
\sin ^{2}\frac{1}{2}\theta & -e^{-i\phi }\cos \frac{1}{2}\theta \sin \frac{1%
}{2}\theta \\
-e^{i\phi }\sin \frac{1}{2}\theta \cos \frac{1}{2}\theta & \cos ^{2}\frac{1}{%
2}\theta
\end{array}
\right] .
\end{equation*}
They may be collected in the general form of a rank-one projector:
\begin{equation}
P(\theta ,\phi )=\frac{1}{2}\left[ \mathbb{I+}\overrightarrow{n}\cdot
\overrightarrow{\sigma }\right] =\frac{1}{2}\left[
\begin{array}{cc}
1+\cos \theta & e^{-i\phi }\sin \theta \\
e^{i\phi }\sin \theta & 1-\cos \theta
\end{array}
\right] ,
\end{equation}
as
\begin{equation*}
P(\theta ,\phi )=\left| m_{+}\theta \phi \right\rangle \left\langle
m_{+}\theta \phi \right| ;\quad P(\pi -\theta ,\pi +\phi )=\left|
m_{-}\theta \phi \right\rangle \left\langle m_{-}\theta \phi \right| .
\end{equation*}
Due to their form, two pairs of eigenvectors $\left| m_{\pm }\theta
\phi \right\rangle ,\left| m_{\pm }\theta ^{\prime }\phi ^{\prime
}\right\rangle $ are not sufficient to yield a basis of projectors,
so that at least three different operators of the family $A(\theta
,\phi )$ are needed to construct a minimal tomographic set, as the
previous example has shown. In the spin case,
starting from the fiducial basis associated with $A(0,0)$%
\begin{equation*}
\left| m_{+}\right\rangle =\left[
\begin{array}{c}
1 \\
0
\end{array}
\right] ,\quad \left| m_{-}\right\rangle =\left[
\begin{array}{c}
0 \\
1
\end{array}
\right] ,
\end{equation*}
any $A(\theta ,\phi )-$basis is related to the fiducial basis \emph{via}%
\textit{\ }the unitary transformation $U(\theta ,\phi ):$%
\begin{equation*}
U(\theta ,\phi )\left| m_{\pm }\right\rangle =\left| m_{\pm }\theta \phi
\right\rangle ,
\end{equation*}
where
\begin{equation}
U(\theta ,\phi ):=\left[
\begin{array}{cc}
e^{-i\phi /2}\cos \theta /2 & e^{-i\phi /2}\sin \theta /2 \\
e^{i\phi /2}\sin \theta /2 & -e^{i\phi /2}\cos \theta /2
\end{array}
\right] .
\end{equation}

Previous analysis has shown that two unitary operators $U(\theta ,\phi
),U(\theta ^{\prime },\phi ^{\prime })$ are a suitably ``skew'' set when $%
\Im (e^{i(\phi ^{\prime }-\phi )}\sin \theta ^{\prime }\sin \theta )\neq 0.$
Thus, in the qu-bit case, a minimal set is obtained by taking a projector
from each pair of eigenvectors and completing the basis with any other of
the remaining projectors. So, as well known \cite{newtonyoung,weigert}, only
three independent directions of $\overrightarrow{n}$\ are needed to
reconstruct a spin 1/2 state, because the fourth projector is along any
direction orthogonal to one of the first three directions.

However, a decomposition of identity involving the whole family
exists and is given by (see the previous eq.s (\ref{decompspinker}),
(\ref{decompspin}))
\begin{equation}
\mathbb{I}=\int_{0}^{2\pi }\int_{0}^{\pi }\hat{K}(\theta ,\phi )\mathrm{Tr}%
(P(\theta ,\phi )\mathbb{\cdot })\sin \theta d\theta d\phi .
\end{equation}
We observe that the kernel $\hat{K}(\theta ,\phi )$ is the simplest,
the only one containing the same few spherical functions which
appear in the projectors $P(\theta ,\phi )$, but it is not unique.
In
fact, a family of equivalent kernels can be obtained by adding to $\hat{K}%
(\theta ,\phi )$\ any other kernel $\hat{K}_{0}(\theta ,\phi )$, containing
only superpositions of spherical functions orthogonal to those of $P(\theta
,\phi ).$

As a matter of fact, in the qu-bit case the equation
\begin{equation}
\mathbb{I}=\int_{0}^{2\pi }\int_{0}^{\pi }P(\theta ,\phi )\mathrm{Tr}(\hat{K}%
(\theta ,\phi )\cdot )\sin \theta d\theta d\phi
\end{equation}
holds in a strong sense. Hence $\hat{K}(\theta ,\phi )$ is just a
Gram-Schmidt kernel in a strong sense.

For the general case of spin $-j\leq m\leq j,$ a decomposition of identity
involving the whole family exists \cite{PhysScr} and reads
\begin{equation*}
\sum\limits_{m=-j}^{j}\int d\Omega \left( R(\theta ,\phi )\left|
m\right\rangle \left\langle m\right| R^{\dagger }(\theta ,\phi )\right)
_{m^{\prime }m^{\prime \prime }}(\hat{K}(m,\theta ,\phi ))_{s^{\prime
}s^{\prime \prime }}=\delta (m^{\prime }-s^{\prime })\delta (m^{\prime
\prime }-s^{\prime \prime })
\end{equation*}
where $R(\theta ,\phi )$ is a rotation through $(\theta ,\phi )$ angles and $%
d\Omega =\sin \theta d\theta d\phi ,$ while
\begin{eqnarray}
(\hat{K}(m,\theta ,\phi ))_{s^{\prime }s^{\prime \prime }}
&=&\sum\limits_{j_{3}=0}^{2j}\sum\limits_{m_{3}=-j_{3}}^{j_{3}}\left(
2j_{3}+1\right) ^{2}\int \left( -1\right) ^{m}D_{0m_{3}}^{\left(
j_{3}\right) }\left( \phi ,\theta ,\gamma \right) \times  \notag \\
&&\left(
\begin{array}{ccc}
j & j & j_{3} \\
m & -m & 0
\end{array}
\right) \left(
\begin{array}{ccc}
j & j & j_{3} \\
s^{\prime } & -s^{\prime \prime } & m_{3}
\end{array}
\right) \frac{d\gamma }{8\pi ^{2}}.
\end{eqnarray}
The problem of a minimal reconstruction formula for spin states was
discussed in Refs. \cite{newtonyoung,weigert}.

\subsection{Photon number tomography}

This is an infinite dimensional case. However, the iso-spectral tomographic
family of operators has a countable discrete spectrum $n=1,2,3,...\infty ,$
so that a generalization of our definitions is straightforward. We assume
the fiducial basis $\left\{ \left| n\right\rangle \right\} $ of the harmonic
oscillator number operator $\hat{a}^{\dagger }\hat{a},$ and with the unitary
family of displacement operators
\begin{equation}
\mathcal{D}\left( \alpha \right) =\exp \left( \alpha \hat{a}^{\dagger
}-\alpha ^{\ast }\hat{a}\right)
\end{equation}
we generate the tomographic family $A\left( \alpha \right) $ depending on
the complex parameter $\alpha :$%
\begin{equation}
A\left( \alpha \right) =\mathcal{D}\left( \alpha \right) \hat{a}^{\dagger }%
\hat{a}\mathcal{D}^{\dagger }\left( \alpha \right) =\left( \hat{a}^{\dagger
}-\alpha ^{\ast }\right) \left( \hat{a}-\alpha \right) ,
\end{equation}
and the respective basis of eigenvectors $\left\{ \left| n\alpha
\right\rangle \right\} =\left\{ \mathcal{D}\left( \alpha \right) \left|
n\right\rangle \right\} $.

The photon number tomogram of a density operator $\hat{\rho}$ is
\begin{equation}
\mathcal{W}_{\rho }(n,\alpha )=\mathrm{Tr}\left( \left| n\alpha
\right\rangle \left\langle n\alpha \right| \hat{\rho}\right)
\end{equation}
while the inversion formula reads
\begin{equation}
\hat{\rho}=\sum\limits_{n=0}^{\infty }\int \frac{d^{2}\alpha }{\pi }\mathcal{%
W}_{\rho }(n,\alpha )K^{\left( s\right) }\left( n,\alpha \right) .
\label{numberinv}
\end{equation}
The operator valued kernel $K^{\left( s\right) }$ is given by
\begin{equation}
K^{\left( s\right) }\left( n,\alpha \right) =\frac{2}{1-s}\left( \frac{s+1}{%
s-1}\right) ^{n}T\left( -\alpha ,-s\right) ,
\end{equation}
where the operator $T$ is
\begin{equation}
T\left( \alpha ,s\right) =\mathcal{D}\left( \alpha \right) \left( \frac{s+1}{%
s-1}\right) ^{\hat{a}^{\dagger }\hat{a}}\mathcal{D}^{\dagger }\left( \alpha
\right) .  \label{opt}
\end{equation}
Here $s$ is a real parameter, $-1<s<1,$ which labels the family of
equivalent kernels $K^{\left( s\right) }\left( n,\alpha \right) .$

The matrix form of eq. (\ref{numberinv}) in the position representation is
\begin{equation*}
\rho \left( x,y\right) =\int dx^{\prime }dy^{\prime }\left[
\sum\limits_{n=0}^{\infty }\int \frac{d^{2}\alpha }{\pi }\left\langle
y^{\prime }\left| n\alpha \right\rangle \left\langle n\alpha \right|
x^{\prime }\right\rangle \left\langle x|K^{\left( s\right) }\left( n,\alpha
\right) |y\right\rangle \right] \rho \left( x^{\prime },y^{\prime }\right)
\end{equation*}
To evaluate the matrix elements of the Gram-Schmidt kernel operator $%
K^{\left( s\right) }$, we first calculate those of the displacement operator
$\mathcal{D}\left( \alpha \right) $. Remembering that $\hat{a}=\left(
Q+iP\right) /\sqrt{2},$ we have
\begin{equation*}
\mathcal{D}\left( \alpha \right) =\exp \left( \left( \alpha -\alpha ^{\ast
}\right) \frac{Q}{\sqrt{2}}-i\left( \alpha +\alpha ^{\ast }\right) \frac{P}{%
\sqrt{2}}\right)
\end{equation*}
and putting
\begin{equation*}
\alpha =\left( \nu -i\mu \right) /\sqrt{2},
\end{equation*}
we get
\begin{equation*}
\left\langle y|\exp \left( -i\mu Q-i\nu P\right) |y^{\prime }\right\rangle
=\delta \left( y-y^{\prime }-\nu \right) \exp \left[ i\left( -\mu y^{\prime
}-\mu \nu /2\right) \right] .
\end{equation*}
Now the following map is useful
\begin{equation*}
\left( \frac{s+1}{s-1}\right) ^{\hat{a}^{\dagger }\hat{a}}=\exp \left[
-i\left( \hat{a}^{\dagger }\hat{a}+1/2\right) \tau _{s}+i\tau _{s}/2\right]
,\quad \tau _{s}:=\left( i\ln \left( \frac{s+1}{s-1}\right) \right)
\end{equation*}
where a determination of $\ln $ has been chosen in such a way that $\tau
_{s}>0$ for $s=0.$ Then one readily obtains
\begin{eqnarray*}
&&\left\langle x|\exp \left[ -i\left( \hat{a}^{\dagger }\hat{a}+1/2\right)
\tau _{s}+i\tau _{s}/2\right] |y\right\rangle \\
&=&\frac{1}{\sqrt{2\pi i\sin \tau _{s}}}\exp \left( i\left[ \left(
x^{2}+y^{2}\right) \cot \tau _{s}-\frac{xy}{\sin \tau _{s}}+\frac{\tau _{s}}{%
2}\right] \right)
\end{eqnarray*}
and the matrix element $\left\langle x|T\left( \alpha ,s\right)
|y\right\rangle $ results as
\begin{equation*}
\int dx^{\prime }dy^{\prime }\left\langle x|\mathcal{D}\left( \alpha \right)
|x^{\prime }\right\rangle \left\langle x^{\prime }\right. |\left( \frac{s+1}{%
s-1}\right) ^{\hat{a}^{\dagger }\hat{a}}|\left. y^{\prime }\right\rangle
\left\langle y^{\prime }|\mathcal{D}\left( -\alpha \right) |y\right\rangle =%
\frac{1}{\sqrt{2\pi i\sin \tau _{s}}}\times
\end{equation*}
\begin{equation*}
\exp \left( i\left[ \left( \left( x-\nu \right) ^{2}+\left( y-\nu \right)
^{2}\right) \cot \tau _{s}-\frac{\left( x-\nu \right) \left( y-\nu \right) }{%
\sin \tau _{s}}-\mu \left( x-y\right) +\frac{\tau _{s}}{2}\right] \right) ,
\end{equation*}
so that eventually the matrix element $\left\langle x|K^{\left( s\right)
}\left( n,\alpha \right) |y\right\rangle $ reads
\begin{equation}
\left\langle x|K^{\left( s\right) }\left( n,\alpha \right) |y\right\rangle =%
\frac{2}{1-s}\left( \frac{s+1}{s-1}\right) ^{n}\frac{1}{\sqrt{2\pi i\sin
\tau _{-s}}}\times
\end{equation}
\begin{equation*}
\exp \left( i\left[ \left( \left( x+\nu \right) ^{2}+\left( y+\nu \right)
^{2}\right) \cot \tau _{-s}-\frac{\left( x+\nu \right) \left( y+\nu \right)
}{\sin \tau _{-s}}+\mu \left( x-y\right) +\frac{\tau _{-s}}{2}\right]
\right) .
\end{equation*}

\subsection{Symplectic tomography}

In the symplectic case ($\hbar =1),$ we start from the fiducial basis $%
\left\{ \left| X\right\rangle \right\} $ of (improper) eigenvectors
of the position operator $Q:Q\left| X\right\rangle =X\left|
X\right\rangle ,$ whose spectrum is the whole real axis: $X\in
\mathbb{R}.$ The two-real parameter
family of unitary canonical operators $S(\mu ,\nu ):$%
\begin{equation*}
S(\mu ,\nu )=\exp i\frac{\lambda }{2}(QP+PQ)\exp i\frac{\theta }{2}%
(Q^{2}+P^{2});\quad \left( \mu =e^{\lambda }\cos \theta ,\nu =e^{-\lambda
}\sin \theta \right) ,
\end{equation*}
generates both an iso-spectral family $A(\mu ,\nu )$ of Hermitian operators
\begin{equation*}
A(\mu ,\nu )=S(\mu ,\nu )QS^{\dagger }(\mu ,\nu )=\mu Q+\nu P
\end{equation*}
and a tomographic set of (improper) eigenvectors $\left| X\mu \nu
\right\rangle =S(\mu ,\nu )\left| X\right\rangle ,$ such that
$\left\langle X^{\prime }\mu \nu |X\mu \nu \right\rangle =\delta
\left( X-X^{\prime }\right) .$ In the position representation
$\{\left| q\right\rangle \}$ it is, for $\nu\ne 0$:
\begin{equation}
\left\langle q|X\mu \nu \right\rangle =\left\langle q|S(\mu ,\nu
)|X\right\rangle =\frac{1}{\sqrt{2\pi |\nu |}}\exp \left[ -i(\frac{\mu }{%
2\nu }q^{2}-\frac{X}{\nu }q)\right] .
\end{equation}

Now, substituting the explicit expression of the symplectic tomogram :
\begin{equation*}
\mathcal{W}_{\rho }(X,\mu ,\nu )=\mathrm{Tr}\left( \left| X\mu \nu
\right\rangle \left\langle X\mu \nu \right| \hat{\rho}\right) =\int
\left\langle q\left| X\mu \nu \right\rangle \left\langle X\mu \nu \right|
q^{\prime }\right\rangle \left\langle q^{\prime }|\hat{\rho}|q\right\rangle
dqdq^{\prime },
\end{equation*}
in the well known inversion formula for $\left\langle
y|\hat{\rho}|y^{\prime }\right\rangle$:
\begin{equation}
\rho (y,y^{\prime })=\frac{1}{2\pi }\int \mathcal{W}_{\rho }(X,\mu ,\nu
)\left\langle y|\exp \left[ i\left( X-\mu Q-\nu P\right) \right] |y^{\prime
}\right\rangle dXd\mu d\nu ,
\end{equation}
we obtain
\begin{eqnarray}
\rho (y,y^{\prime }) &=&\frac{1}{2\pi }\int \left\{ \int \left\langle y|\exp %
\left[ i\left( X-\mu Q-\nu P\right) \right] |y^{\prime }\right\rangle \right.
\notag \\
&&\left. \times \left\langle q^{\prime }\left| X\mu \nu
\right\rangle \left\langle X\mu \nu \right| q\right\rangle dXd\mu
d\nu \right\} \rho (q,q^{\prime })dqdq^{\prime }.
\end{eqnarray}
In different terms
\begin{eqnarray*}
I(y,y^{\prime };q,q^{\prime }) &=&\int \frac{dX}{2\pi }d\mu d\nu
\left\langle y|\exp \left[ i\left( X-\mu Q-\nu P\right) \right]
|y^{\prime }\right\rangle \left\langle q^{\prime }\left| X\mu \nu
\right\rangle \left\langle X\mu
\nu \right| q\right\rangle \\
&=&\delta (q-y)\delta (q^{\prime }-y^{\prime }),
\end{eqnarray*}
so that a partition of identity generated by the the tomographic set of
rank-one projectors $\left| X\mu \nu \right\rangle \left\langle X\mu \nu
\right| $ appears in the inversion formula.

Let us check the previous equation. From
\begin{eqnarray}
&&\left\langle y|\exp \left[ i\left( X-\mu Q-\nu P\right) \right] |y^{\prime
}\right\rangle  \notag \\
&=&e^{iX}\left\langle y|\exp \left( -i\nu P\right) \exp \left( -i\mu
Q\right) \exp \left( -i\mu \nu /2\right) |y^{\prime }\right\rangle  \notag \\
&=&\delta \left( y-y^{\prime }-\nu \right) \exp \left[ i\left( X-\mu
y^{\prime }-\mu \nu /2\right) \right] ,
\end{eqnarray}
we have the following expression of $I(y,y^{\prime };q,q^{\prime }):$%
\begin{equation*}
I(y,y^{\prime };q,q^{\prime })=\int \frac{dXd\mu d\nu }{\left( 2\pi \right)
^{2}\left| \nu \right| }\delta \left( y-y^{\prime }-\nu \right) \times
\end{equation*}
\begin{equation}
\times \exp \left[ i\left( X-\mu y^{\prime }-\mu \nu /2\right) \right] \exp %
\left[ i\frac{\mu }{2\nu }(q^{2}-q^{\prime 2})-i\frac{X}{\nu }(q-q^{\prime })%
\right] .
\end{equation}
Integrating over $X$ we get $\left| \nu \right| \delta \left( q-q^{\prime
}-\nu \right) ,$ which can be used to linearize the quadratic term:
\begin{equation}
\exp \left[ i\frac{\mu }{2\nu }(q^{2}-q^{\prime 2})\right] \delta \left(
q-q^{\prime }-\nu \right) =\exp \left[ i\frac{\mu }{2}(q+q^{\prime })\right]
\delta \left( q-q^{\prime }-\nu \right)
\end{equation}
and we may write $I(y,y^{\prime };q,q^{\prime })$ as
\begin{equation}
\int \frac{d\mu d\nu }{2\pi }\delta \left( y-y^{\prime }-\nu \right) \delta
\left( q-q^{\prime }-\nu \right) \exp \left[ i\frac{\mu }{2}(q+q^{\prime
}-2y^{\prime }-\nu )\right] .
\end{equation}
Integration over $\mu $ yields
\begin{equation}
I(y,y^{\prime };q,q^{\prime })=\int d\nu \delta \left( y-y^{\prime }-\nu
\right) \delta \left( q-q^{\prime }-\nu \right) 2\delta (q+q^{\prime
}-2y^{\prime }-\nu ).
\end{equation}
Eventually, we get the expected result:
\begin{eqnarray}
I(y,y^{\prime };q,q^{\prime }) &=&2\delta \left( q-q^{\prime }-(y-y^{\prime
})\right) \delta (q+q^{\prime }-2y^{\prime }-(y-y^{\prime }))  \notag \\
&=&2\delta \left( q-y-(q^{\prime }-y^{\prime })\right) \delta
(q-y+(q^{\prime }-y^{\prime }))  \notag \\
&=&\delta (q-y)\delta (q^{\prime }-y^{\prime }).
\end{eqnarray}

We conclude this subsection recalling a problem posed by Pauli \cite{pauli},
wether it is possible to recover the state vector of a quantum system from
the marginal probability distributions of the physical observables (e.g.,
position and momentum) of that system. The answer to the question is
obviously negative \cite{reichenbach}. The example of two squeezed states
described by gaussian wave functions,
\begin{equation*}
\psi _{1}\left( x\right) =N\exp \left[ -\alpha x^{2}+i\beta x\right] ,\quad
\psi _{2}\left( x\right) =N\exp \left[ -\alpha ^{\ast }x^{2}+i\beta x\right]
,
\end{equation*}
where $\Re \alpha >0$ and $\beta ^{\ast }=\beta $, demonstrates readily this
negative answer. The moduli of the functions are equal, and also the moduli
of their Fourier transforms are equal. But the states are different since
the scalar product of the two wave functions gives the fidelity which is not
equal to one. So, different wave functions have the same marginal
probability distributions of position and momentum. Now we are able to
understand why the answer must be negative: in fact, the family containing
only the operators position and momentum is not tomographic because it is
too small and cannot generate an inversion formula.

\subsection{Squeeze tomography}

Finally, we discuss an example where an inversion formula is still lacking:
the squeeze tomography \cite{Margav}. The tomogram is defined using the same
unitary operators $S(\mu ,\nu )$ of the symplectic tomography, which acting
on the fiducial basis $\left\{ \left| n\right\rangle \right\} $ of the
photon number tomography generate a basis of squeezed eigenvectors $\left\{
\left| n\mu \nu \right\rangle \right\} =\left\{ S(\mu ,\nu )\left|
n\right\rangle \right\} $ of the squeezed tomographic family $A_{sq}(\mu
,\nu )$%
\begin{equation}
A_{sq}(\mu ,\nu )=S(\mu ,\nu )\hat{a}^{\dagger }\hat{a}S^{\dagger }(\mu ,\nu
)
\end{equation}

In this case, however, the commutant of the family $A_{sq}(\mu ,\nu )$
contains the Parity operator and is nontrivial. Then the family is not a
tomographic family \emph{strictu senso}. Nevertheless, we get a true
tomographic family by a restriction to the subspace of even wave functions.
Then the existence of an inversion formula is granted.

\section{Conclusions}

We summarize the main results of the paper. The mechanism why
quantum states can be described by fair tomographic probabilities
instead of wave functions or density matrices was clarified. The
mathematical reason for the possibility to express the pure state
projector $|\psi \rangle \langle \psi |$ (or density operator
$\hat{\rho}$) in terms of tomograms is based on the simple
observation that any kind of tomogram is just a scalar product of
the projector, treated as a vector in the Hilbert space of
operators, and a basis vector in this Hilbert space. The only
property to be fulfilled is that the basis vectors in that Hilbert
space are a complete (or even an overcomplete) set . For known
examples of tomographies we have shown that it is always so. In view
of this very elementary property it is even mysterious why the
finding of the tomographic probability description of quantum states
was done only relatively recently.

Another result of the paper consists in finding explicit Gram-Schmidt
orthogonalizator kernels for symplectic and photon number tomographies.

Consideration of tomograms of quantum states can be conceptually
extended to the case of many degrees of freedom and even for the
quantum field theory. In fact one needs only a pair of Hilbert
spaces, $\mathcal{H}$ and $B\left( \mathcal{H}\right) ,$ and
constructing a tomographic basis in $B\left( \mathcal{H}\right) $.
In principle, this consideration is extensible to infinite
dimensional case (fields): one only has to add some extra
ingredients to take into account the existence of different
non-unique representations of the infinite Heisenberg-Weyl algebra
and to use extra topological arguments. Another obvious possibility
consists in constructing the basis vectors to provide tomographies
by means of the eigenstates of quantum group operators. For the
$su_q(2)$ case it only needs the introduction of operators dependent
on the Casimir.

\noindent \textbf{Acknowledgements} Vladimir Man'ko, Allen Stern and
George Sudarshan thank University Federico II and INFN, Sezione di
Napoli, for the hospitality extended to them.

\end{document}